\documentclass[12pt,preprint]{aastex}
\usepackage[cmex10]{amsmath}

\shorttitle{Cross Polarised Delay Calibration}
\shortauthors{Cotton}

\begin{document}


\title{A New Method for Cross Polarized Delay Calibration of Radio Interferometers}


\author{W. D. Cotton\altaffilmark{1}}
\affil{National Radio Astronomy Observatory, 520 Edgemont Road, Charlottesville, VA 22903
USA }
\email{bcotton@nrao.edu}


\altaffiltext{1}{The National Radio Astronomy Observatory
(NRAO) is operated by Associated Universities Inc.,
under cooperative agreement with the National Science  Foundation.}


\begin{abstract}
Radio interferometers can measure the full polarization state of
incoming waves by cross--correlating all combinations of two
orthogonal polarizations at each antenna.
The independent sets of electronics used to detect the two
polarization states will introduce a differential instrumental delay
between the two data streams.
The usual technique of separate calibration of the parallel--hand
sets of visibilities still allows for an arbitrary offset in
group delay and phase between the two parallel systems.
In order to use the cross--polarized visibilities, this instrumental
offset must be determined and removed.
This paper describes one such technique and explores its application
in the Obit package.
The technique is successfully applied to some EVLA data using both
strongly and weakly polarized calibrators.
\end{abstract}


\keywords{Data Analysis and Techniques}



\section{Introduction}
Radio interferometry provides a powerful tool for
probing the structure of the polarised emission from celestial
sources.
Such measurements are made by detecting two orthogonal polarization
states at each antenna and cross--correlating all combinations.
To utilize such observations, corrections must be applied for the
effects of the independent electronics in addition to paths through the 
atmosphere and errors in the assumed geometric model that corrupt the
data. 
Most of the geometric and atmospheric effects modifying the arrival
time of a given wavefront are the same for the two polarization states
at a given antenna but differences in the electronics used for the two
polarization states will introduce an instrumental polarized delay and
phase.

The traditional calibration technique for radio
interferometric data, especially when the receptors measure the
orthogonal circular polarizations, has been to determine the
calibration for the two hands of polarization separately.
This effectively ignores both the polarization of the calibrator and
the spurious polarized response of the instrument.

For systems measuring circular polarization and using synchrotron
sources for calibration, this is generally a good approximation.
The circular polarization of the typical compact radio source is less
than 0.1\% and the spurious instrumental polarization response of the
order of a few \%
\footnote{This is discussion is limited to tracking dishes, aperture
arrays such as arrays of dipoles used at low frequencies can have very
large and variable instrumental linear polarizations}.
This approximation is less valid for interferometers measuring
orthogonal linearly polarized signals as the same calibrators
typically have linear polarizations of a percent or less at
frequencies below a few hundred MHz up to $\sim$10\% at frequencies 
above a few tens of GHz \citep{Tucci12}.

Independent calibration of the parallel--hands also has the
disadvantage that it allows an arbitrary group delay and phase offset
between the two parallel systems.
The offsets between the parallel--hands need to be determined and
removed before the cross--polarized visibilities can be used for
astronomical imaging. 
These cross--polarized visibilities may be dominated by the linear
polarization of the source which can be used for cross--polarized
calibration; this case is considered in the following.

One technique used in the past was to ``fringe fit'', fit for group
delay and  delay rate, independently for the parallel--hand
visibilities \citep{Schwab-Cotton-83}, apply this calibration and then
do single baseline delay fitting of a few cross--polarized
visibility spectra and average the results \citep{AIPSMemo79}. 
An alternate technique \citep{Brownetal-89} is to use closure
constraints on the results of a series of single baseline fringe fits
to determine the offsets between the parallel hand systems.
Both of these techniques have limited sensitivity.
This paper explores an improved technique in the Obit package
\citep{OBIT} \footnote{http://www.cv.nrao.edu/$\sim$bcotton/Obit.html}.

\section{Interferometric Polarimetry}
Telescopes using heterodyne electronics as are commonly used in
interferometers at radio wavelengths are sensitive to a single
polarization state of the incoming wave.
In order to fully sample both polarization states of the celestial
signals, sets of electronics nominally sensitive to orthogonal
polarizations are used. 
To sample the full state of the visibility measured with an
interferometer, cross--correlations of all (4) combinations of the
polarizations states are made on each baseline, or pair of antennas.
Denote the various visibilities measured between antennas $j$ and $k$
and using detectors for polarization states $p$ and $q$ as 
$v_{jk}^{pq}$.
Typically, either right-- and left--hand circular polarizations (``R'',
``L'') or orthogonal linear polarizations (``X'', ``Y'') are used in
radio interoferometers. 

In practice, the two polarizations measured are not precisely
those desired but can be modeled by the desired state plus a complex
value, called the ``leakage'' term, $d_{jp}$, times the orthogonal
state.
Thus, the signal received by a detector on antenna $j$ nominally in
polarization state $p$ is actually:
\begin{equation}
s_{jp}' = s_{jp} + d_{jp}s_{jq} 
\end{equation}
The leakage terms will result in a spurious polarized response which
is referred to as ``instrumental polarization'' in the following.

To first order (ignoring products of source and instrumental
polarization terms), the interferometric response for an interferometer
between antennas $j$ and $k$ using linear detectors of an unresolved
source is \citep{TMS2}: 
\begin{equation}\label{XYPol}
\begin{split}
v_{XX}=&{\frac {1}{2}}g_{jX}g_{kX}^*(I+Q{\rm cos}2\chi+U {\rm sin}2\chi), \\
v_{XY}=&{\frac {1}{2}}g_{jX}g_{kY}^*[(d_{jX}+d_{kY}^*)I-
  Q{\rm sin}2\chi+ U {\rm cos}2\chi+iV)], \\
v_{YX}=&{\frac {1}{2}}g_{jY}g_{kX}^*[(d_{YX}+d_{kX}^*)I-
  Q{\rm sin}2\chi+ U {\rm cos}2\chi-iV)], \\
v_{YY}=&{\frac {1}{2}}g_{jY}g_{kY}^*(I-Q{\rm cos}2\chi-U
  {\rm sin}2\chi), 
\end{split}
\end{equation}
where $g_{jp}$ is the complex gain of the electronics for polarization
$p$ on antenna $j$,
``*'' denotes the complex conjugate,
$I$, $Q$, $U$, and $V$ are the Stokes parameters of the source
emission,
$i$ is $\sqrt{-1}$ and,
$\chi$ is the parallactic angle given by:
\begin{equation}\label{ParAng}
\chi\ =\ {\rm tan}^{-1}\Big(\frac{{\rm cos}\ \lambda\ {\rm sin}\ h}
  {{\rm sin}\ \lambda\ {\rm cos}\ \delta\ -\ {\rm cos}\ \lambda\
    {\rm sin}\ \delta\ {\rm cos}\ h}
\Big) 
\end{equation}
where $\delta$ is the source declination, $\lambda$ is the latitude of
the antenna and $h$ is the hour angle of the source.
For linearly polarized detectors with detectors rotated from the
local horizontal and vertical, this rotation needs to be added to the
value of $\chi$ given above.
To simplify the notation in the following, we assume nearly identical
parallactic angles at all antennas at a given time, although the
technique is not limited to this case.

The corresponding first order relations for antennas with circular
detectors is: 
\begin{equation}\label{RLPol}
\begin{split}
v_{RR}\ =\ &{\frac {1}{2}}g_{jR}g_{kR}^*(I\ +\ V), \\
v_{RL}\ =\  &{\frac {1}{2}}g_{jR}g_{kL}^*[(d_{jR}+d_{kL}^*)I\ +\
e^{-2i\chi}(Q\ +\ iU)],\\
v_{LR}\ =\  &{\frac {1}{2}}g_{jL}g_{kR}^*[(d_{jL}+d_{kR}^*)I\ +\
e^{2i\chi}(Q\ -\ iU)],\\
v_{LL}\ =\ &{\frac {1}{2}}g_{jL}g_{kL}^*(I\ -\ V). 
\end{split}
\end{equation}
The impact of source polarization on parallel hand calibration can be
seen from eqs (2) and (4).
Stokes V is generally very small compared to Stokes I, much less than
a percent and can relatively safely be ignored in eq (4).
On the other hand, values of Stokes Q and U of a few percent of Stokes
I will limit the accuracy of parallel hand calibration if ignored in eq
(2).

\cite{SHB96} give a discussion of the three calibration parameters that
cannot be constrained from only parallel hand data.
If the source and instrumental polarizations can be considered
negligible, the third and its frequency derivative can be obtained using
the linear polarization of the calibrator.
For more details on the response of an interferometer to a partially
polarized signal see \citet{TMS2}.

\section {Independent Parallel Hand Calibration}
If the calibrator and instrumental polarizations are sufficiently
small to be ignored, the two parallel polarized sets of visibilities
can be assumed to be independent measurements of the same celestial
quantities. 
For a given baseline, $j-k$, the measured correlations ($v_{jk}^{pp\_obs}$)
are related to the calibrated visibilities  ($v_{jk}^{pp\_cal}$) by:
\begin{equation}
v_{jk}^{pp\_cal}\ = \ v_{jk}^{pp\_obs}\ g_j^p\ g_k^{p*} 
\end{equation}
where $g_j^p$ is given by:
$$g_j^p\ =\ a_j^p\ e^{-2\pi i(\Delta\tau_j^p\nu) - i\phi_j^p}, $$
and $a_j^p$ is the amplitude correction, $\Delta\tau_j^p$ the group
delay residual from the correlator model, $\phi_j^p$ the model phase
residual, and $\nu$ the observing frequency.
Calibration parameters $a_j^p$,  $\Delta\tau_j^p$ and $\phi_j^p$ can
be determined from observations of (calibrator) sources of known
brightness, structure and position. 
Calibration quantities are generally functions of time and frequency. 

This system of equations is degenerate in that only differences in 
$\Delta\tau_j^p$ and $\phi_j^p$ are actually measured.
The system is frequently made determinate by assigning the values of a
``reference antenna'' to zero.
Thus, all phase--like quantities are determined relative to the
reference antenna leaving only the relationship between the two hands of
polarization at the reference antenna undetermined.
If the electronics of the reference antenna are sufficiently stable,
the differences between the two parallel hands at the reference
antenna will be stable in time. 

\section {Residual Cross--hand Group Delay/Phase Offset}
Independent calibration of the parallel--hand systems allows for a
difference between $\Delta\tau_r^p$ and $\Delta\tau_r^q$ as well as between
$\phi_r^p$ and $\phi_r^q$ where $r$ indicates the reference antenna.
Denote these differences as $\Delta\tau_r^{pq}$ and $\phi_r^{pq}$. 
Note: if the observing bandwidth is divided into multiple ``spectral
windows'' with at least partially independent electronics and signal
paths, the calibration parameters may vary between spectral windows.
The relation per spectral window between parallel--hand calibrated data
($v_{jk}^{pq\_cal}$) and cross--hand calibrated data
($v_{jk}^{pq\_xcal}$) is:
\begin{equation}
v_{jk}^{pq\_xcal}\ =\ v_{jk}^{pq\_cal}
e^{-2\pi i\Delta\tau_r^{pq}\nu - i\phi_r^{pq}}.
\end{equation}
Note: this requires a single pair of values,  
$\Delta\tau_r^{pq}$ and $\phi_r^{pq}$, per spectral window.
A sample baseline of EVLA (circular polarization) calibrator data with
parallel--hand but not cross--hand calibration applied is shown in
Figure \ref{noRlDly}. 

\section{A Method of Determining the Cross--hand Offset}
While there are but two parameters to be determined per spectral
window, there are a number of complications.
The most serious of these is spurious instrumental polarization 
($d_{jp}$ in Eqs. \ref{XYPol} and \ref{RLPol})
which can contribute a significant fraction of the strength of the
source polarization to the cross--polarized visibilities.
To first order, this spurious instrumental polarization is independent
of source polarization and can vary strongly with frequency and
baseline. 

A second complication is for arrays with alt-az mounts for which
the antenna rotates with parallactic angle as seem by the source (Eq.
\ref{ParAng}).
If no correction is made and the antennas detect circular
polarization, a varying parallactic angle will cause the source
polarization to rotate in polarization angle while the instrumental
polarization is constant.
If the data are corrected for parallactic angle, the instrumental
polarization then rotates while the source polarization is constant.

A further complication is the case in which the source has
significantly resolved polarized structure.  In this case, the source
component of the cross--polarized response will vary with time,
frequency and baseline.

\subsection{Circularly Polarized Detectorss}
Consider the case of the EVLA with alt-az mounts and detectors sensitive
to circular polarization observing an unresolved source whose linear
polarization is much stronger than the instrumental polarization.
If these data have their phases corrected for parallactic angle and have
parallel--hand calibration applied, the cross--polarized spectra are
from Eq. \ref{RLPol}:
\begin{equation}
\begin{split}
v_{jk}^{RL\_cal}\ &\approx\ (Q + iU)e^{+2\pi i(\Delta\tau_r^{pq}\nu)+i\phi_r^{pq}}\\
v_{jk}^{LR\_cal}\ &\approx\ (Q - iU)e^{-2\pi i(\Delta\tau_r^{pq}\nu)-i\phi_r^{pq}}
\end{split}
\end{equation}
Thus, the $RL$ and $LR$ spectra are the complex conjugates of each
other and this relationship holds for all baselines and times.
In the case represented here, the cross--polarized spectra can be
averaged over all baselines and times and the averaged spectra used to
fit for $\Delta\tau_r^{pq}$ and $\phi_r^{pq}$.

The solution for $\Delta\tau_r^{pq}$ and $\phi_r^{pq}$ can be
separated and done independently.
The group delay ($\Delta\tau_r^{pq}$) can be determined from
the value of $\Delta\tau$ which maximizes:
\begin{equation}\label{DelayFit}
 |\Sigma(v_{jk}^{RL\_cal} +
{{v_{jk}^{LR\_cal*}})e^{-2\pi i(\Delta\tau\nu)}}|
\end{equation}
using a direct parameter search over $\Delta\tau$ and $||$ denotes the
modulus.
The summation is over spectra for all included times and baselines.

The phase ($\phi_r^{pq}$) is then the phase of the average of the
delay corrected visibilities: 
\begin{equation}\label{PhaseFit}
\phi_r^{pq}\ =\  arg[\Sigma(v_{jk}^{RL\_cal} +
{v_{jk}^{LR\_cal*}})e^{-2\pi i(\Delta\tau_r^{pq}\nu)}] 
\end{equation}
The analysis above is only valid if the cross--polarised visibilities
include a significant detection of the polarized signal.
The typical ``signal--to-noise ratio'' ($SNR$) of the individual
channel samples can be estimated from the RMS scatter of their
phases about the mean in radians and the approximation that:
\begin{equation}\label{SNRFit}
SNR\ \approx\ {\frac{1}{RMS_{phase}}}
\end{equation}
for an $SNR\ >$ 1.
Channel averages over baseline and time can be used to increase the
$SNR$  of individual samples.

In general, the instrumental polarization is not negligible compared
to the source polarization but should vary from baseline-to-baseline.
Furthermore, after the data are corrected for the parallactic angle,
the instrumental polarization will vary with parallactic angle, hence
time.
Averaging over baselines and time will reduce the contribution of the
instrumental polarization to the averaged $RL$ and $LR$ spectra.

\subsection{Linearly Polarized Detectors}
The case of interferometers, such as ALMA or the ATCA, whose elements
are sensitive to linear polarization, is more complex.
As can be see from Eq. \ref{XYPol},
the source polarization component of the cross--hand visibility
spectrum is: 
\begin{equation}
\begin{split}
v_{jk}^{XY\_cal}\ &\approx\ (-Q{\rm sin}2\chi+ U {\rm cos}2\chi+ iV)
e^{+2\pi i(\Delta\tau_r^{pq}\nu)+i\phi_r^{pq}}\\
v_{jk}^{YX\_cal}\ &\approx\ (-Q{\rm sin}2\chi+ U {\rm cos}2\chi- iV)
e^{-2\pi i(\Delta\tau_r^{pq}\nu)-i\phi_r^{pq}}.
\end{split}
\end{equation}
Stokes V for most sources is very small and can be ignored.
However, the linearly polarized component is a more complex function
of parallactic angle and can only be averaged over a time range and
set of baselines for which the variation of parallactic angle is
small.

The averaged XY and YX spectra can be used to determine the X-Y delay
difference as was done for circular feeds in the section above and the
SNR of the data can be determined as well.
Variation of the instrumental polarization among baselines will
reduce its effect on the averaged spectra.

\section{RLDly: An Obit Implementation}
The technique outlined above has been implemented in the
Obit\citep{OBIT} package available through task {\tt RLDly}.
The fitting is that described in the previous sections except using
weighted sums to fit for $\Delta\tau_r^{pq}$ and $\phi_r^{pq}$.
The weights used are the data quality weights assigned to the data
which are inversely proportional to the visibility noise
variance.
The implementation in {\tt RLDly} allows for dropping the end channels in
each spectral window which may be poor representations of the values
being fitted.
In this case the phase ramp applied to the spectra still begins with
the first channel in each spectrum.

\section{Testing}
The calibration data from an EVLA observation was used to test this
technique.
The data included the bandpass 6.0--8.0 GHz in 16 spectral windows
(IFs) each averaged to 16 channels.
The observations included 5 $\times$ 15 second scans over roughly 5
hours on the strongly linearly polarized calibrator 3C286 ($\approx10$ \%).
The EVLA has detectors sensitive to circular polarization and alt-az
antenna mounts.

\begin{figure*}
\centerline{
  \includegraphics[angle=-90,width=7.0in]{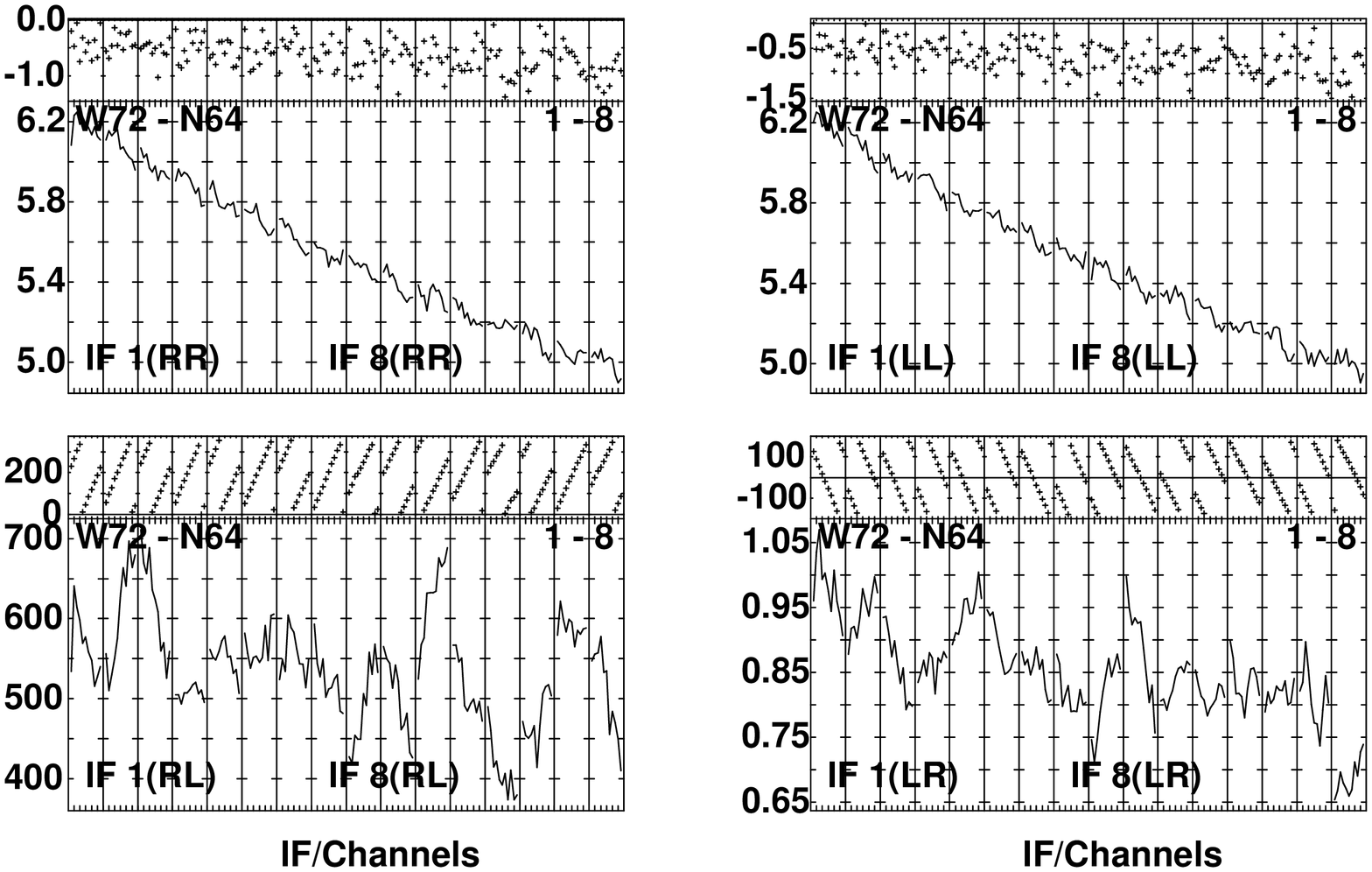}
}
\caption{ 
Sample EVLA visibility spectra for 3C286 averaged over a 15 second
scan, parallel--hand visibities above, cross--hand below.
Parallel--hand calibration has been applied but not the differences
between the R and L systems.
The upper plot in each panel gives the phase in degrees and the lower,
the amplitude in Jy or mJy. 
Note the rapid variation of cross--polarized phase with frequency
indicating a large delay offset.
} 
\label{noRlDly}
\end{figure*}

Data had strong, non--celestial signals removed and were initially
calibrated using the standard parallel--hand 
techniques; a sample of these data on one of the longer baselines is
shown in Figure \ref{noRlDly}. 
Corrections were applied for parallactic angle so all times were used
in the fitting.

The data were obtained in the ``A'' configuration for which the
synthesized beam is about 0.25'' FWHM.
3C286 is well resolved in this configuration but the bulk of the
emission, both total and polarized intensity is from the marginally
resolved core.
Therefore, all baselines were used for the fitting.
A target R-L phase of 66$^\circ$ and rotation measure of 0 rad/m were
provided. 
The outer 3 channels on each end of each spectral window were omitted
from the fitting of the 3C286 data.

\begin{figure*}
\centerline{
  \includegraphics[angle=-90,width=7.0in]{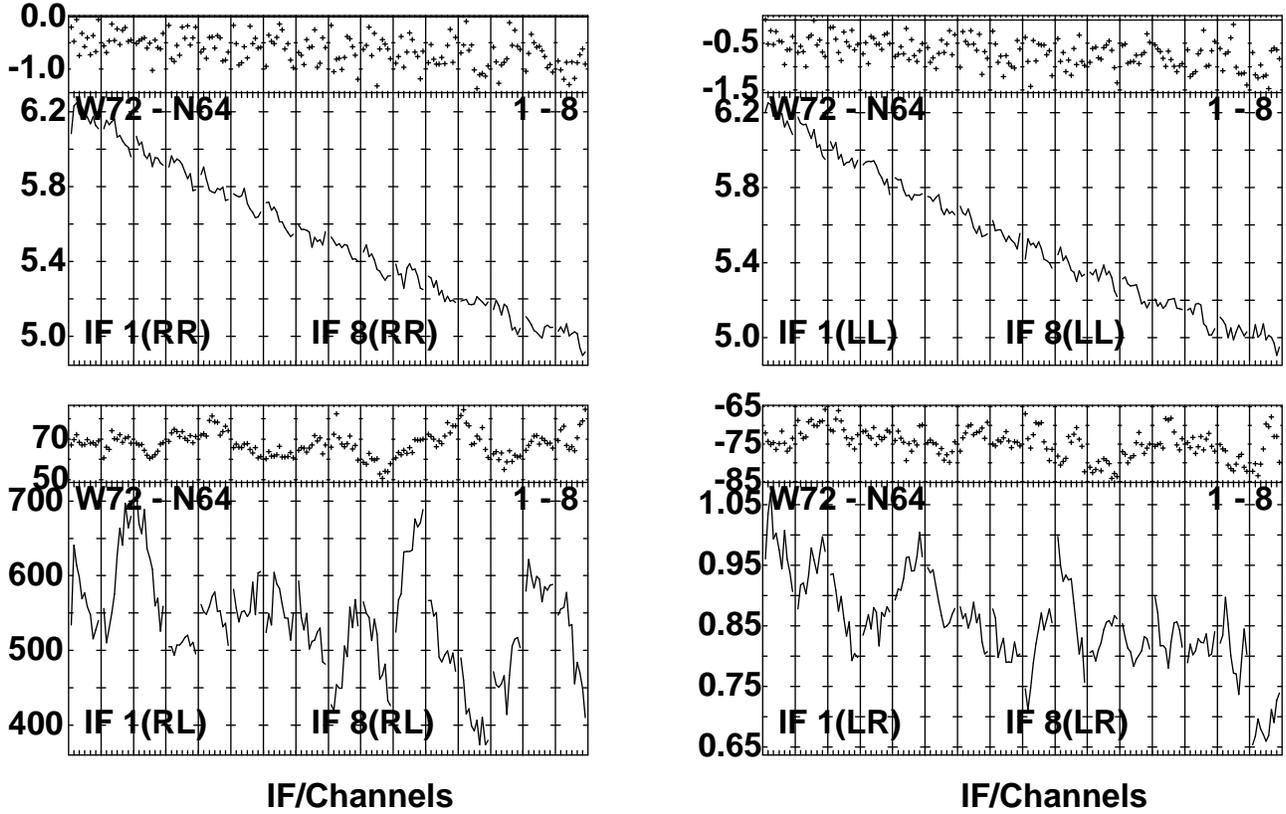}
}
\caption{ 
Data shown in Figure \ref{noRlDly} but with the R--L delay and phase
offsets applied.
The strongly polarized source 3C286 (data shown) was used for the
calibration. 
Note: the parallel--hand data is unaffected.
Instrumental polarization corrections have not been applied.
} 
\label{wRlDly}
\end{figure*}

The cross--polarized corrections were applied to the data.
The same data shown in  Figure \ref{noRlDly} after this recalibration
are shown in Figure \ref{wRlDly}.
The rapid variation of the cross--polarized phase with frequency is
largely eliminated.
Variations of phase with frequency remaining are due to the frequency
variable instrumental polarization which has not been corrected;
these can be corrected using AIPS task PCAL.
The fluctuations of instrumental polarization are also visible in the
cross--polarized amplitudes.
Note that the parallel--polarized data are unaffected by the
cross--polarized calibration.

\begin{figure*}
\centerline{
  \includegraphics[angle=-90,width=7.0in]{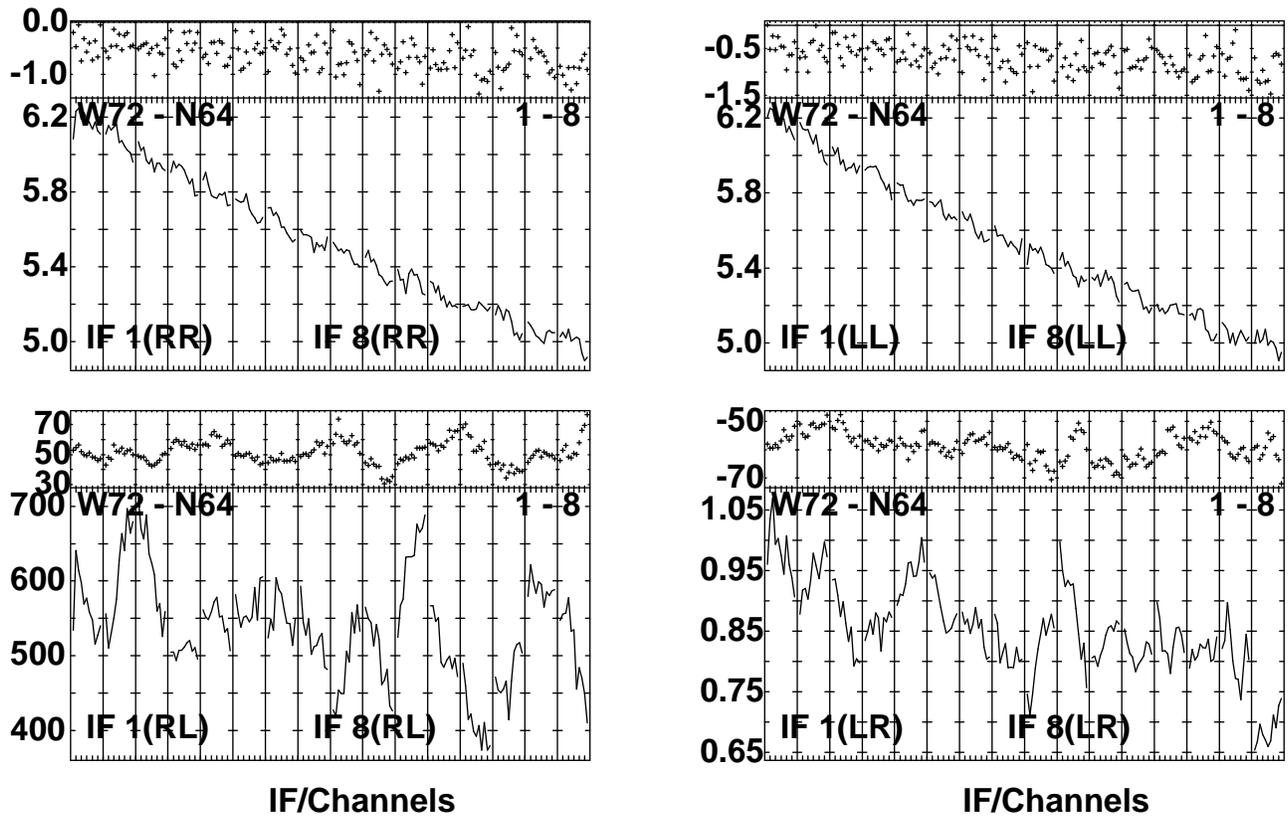}
}
\caption{ 
Data shown in Figure \ref{wRlDly} but with the R--L delay determined
from the weakly polarized calibrator J1504+1029.
} 
\label{wRlDly1504}
\end{figure*}

In order to test the effect of a weakly polarized calibrator, the same
test was run but using as calibrator J1504+1029 whose polarization
($\approx1$ \%) is less than the typical baseline instrumental polarization.
A plot of the data shown in Figures \ref{noRlDly} and \ref{wRlDly} 
after calibration using this source is given in Figure \ref{wRlDly1504}.

\section{Discussion}
A relatively general and robust method for fitting instrumental
cross--polarized group delays and phases is presented and tested using
EVLA data. 
In these tests, the large phase slope in the cross--polarized spectra
due to the difference in the parallel--hand group delays is essentially
removed. 

The fitting of data from circularly polarized detectors and a strongly
linearly polarized unresolved calibrator, as in the first test
presented, is relatively straightforward as all baselines are
measuring essentially the same value and much averaging can be
incorporated. 
More extensive testing will be needed in the case of 
weakly polarized calibrators where the cross-polarized visibilities
will be dominated by instrumental polarization and arrays using
linearly polarized detectors for which time averaging need be more
limited. 
One test on the same dataset as discussed above using a calibrator whose
polarized emission is less than the typical baseline instrumental
polarization resulted in cross--polarized delay fits comparable to
those derived from 3C286 although with larger scatter in the R--L phase.

Fitting for cross--polarized group delays is needed prior to fitting for
instrumental polarization if any averaging in frequency is to be used
in the calibration fitting.
Fitting for cross--polarized phases at this point will be only
approximate due to corruption by the instrumental polarization and
will need to be redetermined after removal of the instrumental
polarization. 

The technique described here can also be applied to VLBI data for
which the parallactic angle is different amoung the various antennas at a
given time if the data are initially corrected for parallactic angle
(circular feeds are generally used for VLBI).
At VLBI resolutions there is generally significant resolution of the
polarized emission of even calibrators and a limited range of
baselines may need to be used.

\acknowledgments

The author would like to thank the anonymous reviewer for many helpful
comments, corrections and suggestions that led to a significant
improvement in the paper. 

{\it Facilities:} \facility{VLA}.


\clearpage

\begin{thebibliography}{}
\bibitem[Brown, Roberts \& Wardle (1989)]{Brownetal-89} 
Brown, L. F., Roberts, D. H. \& Wardle, J. F. C, 1989, \aj, 87, 1522
\bibitem[Cotton (2008)]{OBIT} 
Cotton, W. D., 2008, \pasp, 120, 439
\bibitem[Cotton (1992)]{AIPSMemo79} 
Cotton, W. D, 1992, AIPS Memo Series, 79, 1 \hfil\break
ftp://ftp.aoc.nrao.edu/pub/software/aips/TEXT/PUBL/AIPSMEMO79.PS
%
\bibitem[Sault, Hamaker \& Bregman (1996)]{SHB96} 
{Sault}, R.~J. and {Hamaker}, J.~P. and {Bregman}, J.~D., 1996,
  \aaps,  117, 149
\bibitem[Schwab \& Cotton (1983)]{Schwab-Cotton-83}  
Schwab, F. R. \&  Cotton, W. D., 1983, \aj, 88, 688
\bibitem[Thompson, Moran \& Swenson (2001)]{TMS2} 
Thompson, A.~R., Moran, J.~M. \& Swenson, G.~W. Jr., 2001,
{\bf Interferometry and Synthesis in Radio Astronomy}, 2nd Edition,
Wiley-Interscience
\bibitem[Tucci \& Toffolatti (2012)]{Tucci12}  
Tucci, M. \&  Toffolatti, L., 2012, arXiv:1204.0427v1[astro-ph.CO]
\end{thebibliography}
\end{document}